\documentclass[traditabstract]{aa}

\usepackage{graphicx}
\usepackage{natbib}
\bibpunct{(}{)}{;}{a}{}{,}

\newcommand{\adc}{2A~1822--371}

\newcommand{\xmm}{\it XMM-Newton}
\def\ltsima{$\; \buildrel < \over \sim \;$}
\def\simlt{\lower.5ex\hbox{\ltsima}}
\def\gtsima{$\; \buildrel > \over \sim \;$}
\def\simgt{\lower.5ex\hbox{\gtsima}}

\begin{document}

\title{New ephemeris of the ADC source {\adc}:
A stable orbital-period derivative over 30 years}

\author{L.~Burderi$^{1}$\thanks{mail to burderi@dsf.unica.it} \and 
T.~Di Salvo$^{2}$\thanks{mail to disalvo@fisica.unipa.it} \and 
A.~Riggio$^{1,3}$ \and A.~Papitto$^{1,3}$ \and R.~Iaria$^{2}$ \and 
A.~D'A\`i$^{2}$ \and M.~T.~Menna$^{4}$ 
}					 

\institute{Dipartimento di Fisica, Universit\`a degli Studi di Cagliari, 
SP Monserrato-Sestu, KM 0.7, Monserrato, 09042 Italy
\and Dipartimento di Scienze Fisiche ed Astronomiche, Universit\`a di Palermo, 
via Archirafi 36, Palermo, 90123, Italy 
\and INAF Osservatorio Astronomico di Cagliari, Loc. Poggio dei Pini, 
Strada 54, 09012 Capoterra (CA), Italy
\and INAF Osservatorio Astronomico di Roma, via Frascati 33, Monteporzio 
Catone, 00040, Italy  }

\abstract
{We report on a timing of the eclipse arrival times of the low mass 
X-ray binary and X-ray pulsar {\adc} performed using all  
available observations of the Proportional Counter Array on 
board the {\it Rossi} X-ray Timing Explorer, {\xmm} pn, and Chandra.
These observations span the years from 1996 to 2008.
Combining these eclipse arrival time measurements with those already
available covering the period from 1977 to 1996,  
we obtain an orbital solution valid for more than thirty
years.
The time delays calculated with respect to a constant orbital 
period model show a clear parabolic trend, implying that the orbital 
period in this source constantly increases with time at a rate
$\dot P_{orb} = 1.50(7) \times 10^{-10}$ s/s. This is 3 orders of 
magnitude larger than what is expected from conservative mass transfer 
driven by magnetic braking and gravitational radiation. 
From the conservation 
of the angular momentum of the system we find that to explain 
the high and positive value of the orbital period derivative 
the mass transfer rate must not be less than 3 times the Eddington 
limit for a neutron star, suggesting that the mass transfer has to be 
partially non-conservative.
Under the hypothesis that the neutron star accretes at the Eddington limit
we find a consistent solution in which at least 70\% of the transferred 
mass has to be expelled from the system.}

\titlerunning{A Stable Orbital Period Derivative over the Last 30 Years}
\authorrunning{L. Burderi et al.}

\maketitle

\section{Introduction}

The source {\adc} is a well-known low mass X-ray binary (hereafter LMXB) 
seen almost edge-on with an inclination angle of $i \sim 85^\circ$ 
\citep{Hellier_89}. The observed average unabsorbed flux of the source 
in the $0.1-100$ keV energy range is $\sim 1.5 \times 10^{-9}$ ergs 
cm$^{-2}$ s$^{-1}$ \citep{Iaria_01}. 
This corresponds to an unabsorbed luminosity of $\sim 1.2 \times 10^{36}$ 
ergs s$^{-1}$, adopting a distance of 2.5 kpc \citep{Mason_82}. However, 
it has been noted \citep{Parmar_00} that the mean ratio of the X-ray
over optical luminosity, $L_X / L_{opt}$, for {\adc} is about 20, while 
the average value for LMXBs is about 500 \citep{vanParadijs_94}.
This would imply an unobscured X-ray luminosity as high as 
$3 \times 10^{37}$ ergs/s for the assumed distance of 2.5 kpc.
The apparent low luminosity of the source has therefore to be ascribed 
to the high inclination of the system with respect to the 
line of sight.
Indeed, the light curve of {\adc} shows both dips and eclipses of the 
X-ray source by the companion star. The 
partial nature of the eclipse indicates that the X-ray emitting region is 
extended and that the observed X-rays are scattered in an accretion disk 
corona (ADC, \citet{White_81}).  The X-ray light curve shows clear signs 
of orbital modulation with a binary orbital period of $5.57$ h.
This X-ray modulation is probably caused by the obscuration of the
ADC by the thick rim of an accretion disk. The orbital period has been 
measured from eclipse timing to increase gradually \citep{Hellier_90}.
\citet{Parmar_00} gave the best ephemeris of this source before this 
work. In particular they found a significant positive orbital period 
derivative of $\dot P_{\rm orb} = 1.78 \times 10^{-10}$ s/s.

\citet{Jonker_01} reported on the discovery of 0.59~s X-ray pulsations
in this source in an RXTE observation performed in 1998. The timing analysis
of the pulse arrival times indicates a circular orbit with an eccentricity 
$e < 0.03$ (95\% c.l.) and an $a \sin i$ for the neutron star of 
1.006(5)~lt-s, implying a mass function of $(2.03 \pm 0.03) \times 10^{-2}\; 
M_\odot$.  The comparison between the pulse period measured by RXTE in 1996
and 1998 also indicates that the neutron star in this system 
is spinning up at a rate of $\dot P = (-2.85 \pm 0.04) \times 10^{-12}$ s/s.
\citet{Jonker_01} inferred a bolometric X-ray luminosity of about 
$(2-4) \times 10^{37}$ ergs/s assuming a magnetic field of $(1-5) \times 
10^{12}$ Gauss. From spectroscopic measurements of the radial velocity curve 
of the companion, \citet{Jonker_03} derived a lower limit to the mass of 
the neutron star and to that of the companion star of $0.97 \pm 0.24$ and 
$0.33 \pm 0.05\; M_\odot$, respectively (1~$\sigma$, including uncertainties 
in the inclination), and an accurate estimate of the system inclination 
angle, $i= 82^\circ.5$.

In this paper we report on the analysis of X-ray observations of {\adc} 
performed from 1996 to 2008 by RXTE, {\xmm}, and Chandra with the aim to 
derive eclipse arrival times and to improve the orbital ephemeris. We
confirm with higher precision and over a much larger time span (about 31
years) the ephemeris found by \citet{Parmar_00}. In particular we find 
that the orbital period derivative has remained constant during the last 
30 years. Finally we discuss the implications of a high and positive value of 
the orbital period derivative on the mass transfer rate and secular evolution 
of this source.

\begin{table*}[th]
\caption{New X-ray eclipse times for {\adc}.  }
\label{tabobs}
\centering
\renewcommand{\footnoterule}{}  
\begin{tabular}{lllll}
\hline \hline
T$_{ecl}$ (MJD)  & Error & Cycle & Satellite (ObsID) 
& Tstart - Tstop (MJD) \\ 
\hline
50352.10094 & 0.00047 & 20410 &  RXTE (P10115) & 50352.34 - 50353.44 \\
50992.0857  & 0.0023  & 23167 &  RXTE (P30060) & 50992.81 - 51019.65 \\
51975.01132 & 0.00031 & 27402 &  RXTE (P50048) & 51975.44 - 52101.48 \\
52432.03655 & 0.00030 & 29371 &  RXTE (P70036/37) & 52432.37 - 52435.77 \\
52487.97497 & 0.00038 & 29612 &  RXTE (P70037) & 52488.36 - 52504.01 \\
52519.07766 & 0.00085 & 29746 &  RXTE (P70037) & 52519.15 - 52547.87 \\
52882.09667 & 0.00037 & 31310 &  RXTE (P70037) & 52882.02 - 52885.21 \\
51975.06934 & 0.00056 & 27402 & XMM-Newton (230101) & 51975.55 - 51976.15 \\
51779.6317  & 0.0019  & 26560 & Chandra (671) & 51779.70 - 51780.16 \\
54607.19592 & 0.00056 & 38742 & Chandra (9076/9858) & 54606.96 - 54610.52 \\
\hline
\end{tabular} 
\vskip 0.5cm
Note: Uncertainties are calculated as described in the text.
Tstart - Tstop indicates the time interval over which a folding of 
the orbital light curve was performed to derive the 
time of eclipse.
\end{table*}

\section{Timing analysis and results}

We analysed all available X-ray observations of {\adc} 
performed over the period from 1996 to 2008. In particular we used 
observations from the PCA on board RXTE performed in 1996 (P10115), 
1998 (P30060), 2001 (P50048), 
2002 (P70036), 2002-2003 (P70037), 
one observation from {\xmm} performed in 2001 (Obs ID: 0111230101 and
0111230201), and two Chandra observations performed in 2000 
(Obs ID: 671) and in 2008 (Obs ID: 9076 and 9858), respectively.
The arrival times of all events were referred to the solar system 
barycenter, using as the best estimate for the source coordinates 
those derived from the 2008 Chandra observations (RA: 18 25 46.81,
DEC: -37 06 18.5, uncertainty: $0.6''$).

The typical eclipse duration is around 2.2 ks, which corresponds to
$10\%$ of the binary orbital period. 
In order to improve the statistics for the measure of the 
eclipse epochs and to have the possibility of fitting a complete 
orbital light curve we decided to perform a folding of these data
using the known binary orbital period of the source,
after verifying that this folding does not affect the results reported 
here in any case. Folding the data is not an important 
issue for the two Chandra observations and the XMM observation, 
where just one or two consecutive eclipses are observed. 
But it is important for the RXTE observations, because
these are short and sparse, and also because the RXTE observations
are continuously interrupted by the Earth occultation at every RXTE
orbit (lasting approximately 1.5 h). In this case the folding is
required to sample a complete orbital light curve from the 
source, because this is important for a meaningful fitting of the eclipse.
For each of these observations we hence folded the data using the local 
orbital period as derived from the ephemeris published by 
\citet{Parmar_00}. The 2002-2003 RXTE dataset (P70036 and P70037) 
was long enough and we decided to divide it into the following 
four periods: i) 2002 June 7-10, ii) 2002 August 2-18, 
iii) 2002 September 2-30, and iv) 2003 August 31 - September 3. 
In this way we obtained a total of 10 orbital light curves
in which the eclipses were clearly visible (see Table~\ref{tabobs} for 
details on the used observations). 

We then fitted these orbital 
light curves to derive eclipse arrival times with the 
procedure described below. Because the eclipses are 
asymmetrical and partial, the exact eclipse centroid times crucially 
depend on the model adopted to describe their shape as well as the 
variable continuum they are superimposed on. In order to be 
conservative in our estimates, we then decided to fit the
folded light curves using 10 different models.
The first model is that used by \citet{Parmar_00} 
consisting of a Gaussian and a constant fitted on a phase 
interval of 0.1 around the eclipse. The second and third models 
consist again of a Gaussian and a constant plus a linear term (second model)
and a linear and quadratic term (third model) fitted on a phase interval 
of 0.3 around the eclipse. The fourth model is as the third model plus
a cubic term fitted on a phase interval of 0.4 around the eclipse.
The fifth model consists of a Gaussian and a constant plus a sinusoid
of period fixed to the orbital period fitted on the whole 0-1 phase
interval. The models from the sixth one to the tenth one are as the fifth model, 
plus from 2 to 6 sinusoids with periods fixed to 1/2 up to 1/6 of the orbital 
period, respectively. The addition of higher harmonic components was required 
to better describe the overall orbital light curve shape,
which differs from a pure sinusoid. We restricted our fitting to the
first six harmonics because the addition of higher harmonic components was 
not statistically significant based on an F-test.

Thus we obtained 10 eclipse arrival times (each corresponding to one of 
the models described above) for each orbital light curve. The final 
eclipse arrival time for each orbital light curve was chosen to be
the average of these 10 values, and the associated uncertainty was 
chosen to be half of the maximum range spanned by these values 
($1\; \sigma$ error included). The uncertainty derived in this way
fully takes into account significant discrepancies among the different 
eclipse arrival times found with a particular model to describe the 
eclipse and the orbital modulation. 
The obtained values of the eclipse epochs for each of the 10
orbital light curves and the relative uncertainties are reported in 
Table~\ref{tabobs}.

\begin{figure}
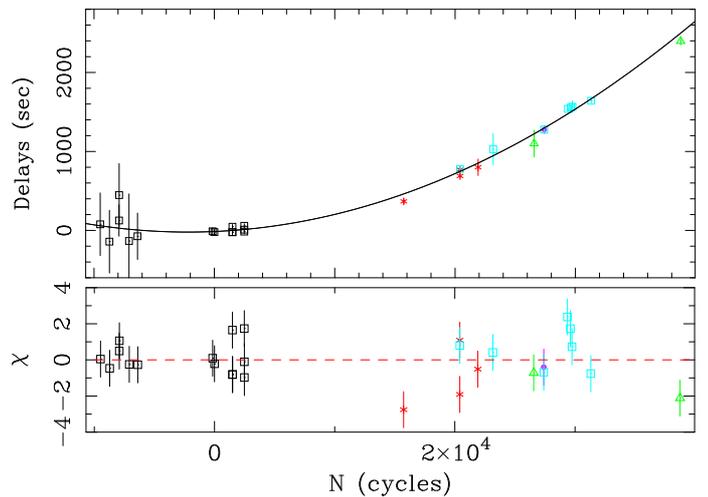

\includegraphics[angle=-90,width=9.0cm]{12881fg1a.ps}
\includegraphics[angle=-90,width=9.0cm]{12881fg1b.ps}
\caption{Eclipse time delays with respect to a constant orbital
period model plotted vs.\ the orbital cycle for all the available
eclipse time measures spanning the period from 1977 to 2008 
together with the best-fit parabola (top panel), and residuals in
units of $\sigma$ with respect to the best-fit parabola (bottom panel).
Different symbols indicate different datasets: black squares are from 
\citet{Hellier_90}, red stars are from \citet{Parmar_00}, cyan
squares are from RXTE data, the magenta dot is from 
{\xmm} data, and the green triangles are from Chandra data.}
\label{fig:timedelays}
\end{figure}

We then computed the eclipse time delays by subtracting from our 
measures the eclipse arrival times predicted by a constant orbital
period model adopting the orbital period, $P_{\rm orb \; 0}$, and the 
reference time, $T^e_0$, given by \citet{Parmar_00}. 
These time delays were plotted versus the orbital cycle number $N$. 
The integer $N$ is the exact number of orbital cycles elapsed since 
$T^e_0$; i.e., $N$ is the closest integer to 
$(T^e_N - T^e_0)/ P_{\rm orb \;0}$ under the assumption that 
$| T^e_N - (T^e_0 + N P_{\rm orb \; 0}) | << P_{\rm orb \; 0}$ that we 
have verified {\it a posteriori}.
These results are shown in Fig.~\ref{fig:timedelays} 
together with all delays computed from previously
available eclipse times, namely those given by \citet{Hellier_90} 
and by \citet{Parmar_00}, respectively.

These points show a clear parabolic trend that we fitted to the equation
\begin{equation}
\delta T^e_N = \delta T^e_0 + \delta P_{\rm orb \; 0} \times N + 
(1/2) \dot P_{\rm orb} P_{\rm orb \; 0} \times N^2,
\label{eq1}
\end{equation}
where the correction to the adopted value of the eclipse time, 
$\delta T^e_0$, and to the adopted value of the orbital period, 
$\delta P_{\rm orb \; 0}$, and the orbital period derivative, 
$\dot P_{\rm orb}$, are the fit parameters.
We get a very good fit with a $\chi^2 / d.o.f. = 38.69 / 25 = 1.5$.
In agreement with previous results, we find a highly significant derivative 
of the orbital period, which indicates that the orbital period in this 
system is increasing at a rate of $\dot  P_{\rm orb} = (1.499 \pm 0.071) 
\times 10^{-10}$ s/s. 
The best-fit values for the orbital parameters, calculated with the
corrections we found from the fit of the parabolic trend of the eclipse epochs 
with Eq.~\ref{eq1}, are shown in Table~\ref{tabps}. 
Note that a similar orbital period derivative was recently found
with new measures of optical eclipses by \citet{Bayless_09}.

\begin{table}
\begin{minipage}[t]{\columnwidth}
\caption{Best-fit orbital solution for {\adc} derived from the analysis of
the eclipse arrival times from 1977 to 2008.}
\label{tabps}
\centering
\renewcommand{\footnoterule}{}  
\begin{tabular}{llrr}
\hline \hline
Parameter  & Units & P2000 & This work  \\
\hline
$T^e_0$ & MJD  	& $45614.80964(15)$ & $45614.80948(14)$ \\
$P_{\rm orb \; 0}$  & s	& $20054.1990(43)$ & $20054.2056(22)$ \\
$\dot P_{orb}$ & $ 10^{-10}$ s/s & $1.78(20)$ & $1.499(71)$ \\
$\chi^2 / d.o.f.$ &              & $21.4 / 16$  & $38.69 / 25$ \\
\hline
\end{tabular} 
\end{minipage}
Note: 
Errors are at $1~\sigma$ c.l. on the last 2 digits. 
The value of $P_{\rm orb \; 0}$ is referred to $T^e_0$. The best-fit
orbital parameters reported in this work are compared with the 
ephemeris given by P2000 \citep{Parmar_00}.
\end{table}

\section{Orbital evolution of {\adc}}


Apart from mass transfer between the companion and the neutron star,
the orbital evolution of this binary system is expected to be driven
by the emission of gravitational waves and by magnetic braking. Under
the further assumption of conservative mass transfer, orbital
evolution calculations show that the orbital period derivative should
be
$$
\dot P_{\rm orb} =  -3.0 \times 10^{-14} \; m_1 \; m_{2, \; 0.1} \; m^{-1/3} 
P_{\rm 5h}^{-5/3} \times [1.0 + {\rm T_{MB}}]  
$$
\begin{equation} 
\times \; [(n - 1/3) / (n + 5/3 - 0.2 m_{2, \; 0.1} m_1^{-1})] \; {\rm s \; s^{-1}} 
\end{equation}
(see \citet{Disalvo_08, Verbunt_93}; see also \citet{Rappaport_87}),
where $m_1$ and $m$ are the mass of the primary, $M_1$, and 
the total mass, $M_1 + M_2$, in units of $M_\odot$ respectively, $m_{2, \; 0.1}$ is 
the mass of the secondary in units of $0.1\; M_\odot$, $P_{\rm 5h}$ is the 
orbital period in units of 5~h (that is appropriate for {\adc} because 
$P_{\rm orb} = 5.57$ h),  
$n$ is the index of the mass-radius relation of the secondary 
$R_2 \propto M_2^{\rm n}$, 
and where the term ${\rm T_{MB}} \sim 20$ takes into account the effect 
of the magnetic braking.

In line with \citet{Verbunt_81}, \citet{Verbunt_93}, and \citet{King_88}
(see \citet{Tauris_01} for a review), we can parametrise this term as
\begin{equation}
\label{eq:MB}
{\rm T_{MB}} = 19.3 \; (f/k_{0.277})^{-2} m_{2, \; 0.1}^{1/3} m_1^{-4/3} P_{5h}^{2},
\end{equation}
where $f$ is a dimensionless parameter of order of unity: preferred values are
$f = 0.73$ \citep{Skumanich_72} or $f = 1.78$ \citep{Smith_79}, and $k_{0.277}$
is the radius of gyration of the star $k$ in units of $0.277$, which is the 
appropriate value for a $1 \; M_\odot$ ZAMS star \citep{Claret_89}. 
Note that the expression for the MB term given in \citet{Verbunt_93} is 
recovered from the above adopting $f = 1$ and $k$ for a 
$1 \; M_\odot$ ZAMS star. Actually, \citet{Tauris_01} discussed three different
expression for the angular momentum losses due to MB, $\dot J_{\rm MB}$, 
namely that proposed by Skumanich \citep{Verbunt_81}, that proposed by
\citet{Stepien_95}, and that proposed by \citet{Rappaport_83}, respectively.
However $|\dot J_{\rm Stepien}| \le 0.1 |\dot J_{\rm Skumanich}|$ with 
$|\dot J_{\rm Rappaport}|$ somewhat between them. Because we found that
to describe {\adc} a quite large $\dot J_{\rm MB}$ is required, 
we decided to adopt $|\dot J_{\rm Skumanich}|$ which resulted in the 
term ${\rm T_{MB}} \sim 20$ adopted above.

The orbital period derivative we measured cannot be explained
by a conservative scenario however.
A positive orbital period derivative certainly indicates a 
mass-radius index $n < 1/3$; this is indeed a quite general result,
which does not depend on the details of the angular momentum losses 
(see also Eq.~\ref{eq:dotM} below). 
However, the orbital period derivative we measured, 
$\dot P_{\rm orb} = 1.50(7) \times 10^{-10}$ s/s, is about three 
orders of magnitude 
larger than what is expected even including the (strongest) MB term!
This discrepancy is embarrassingly large suggesting that the
conservative evolutionary scenario cannot be applied in this case.
A similar conclusion was reached by \citet{Bayless_09}, who
give an improved ephemeris for this source based on new optical eclipse
measures; these authors also note that an extremely high mass accretion
rate onto the neutron star, corresponding to about four times the Eddington 
limit, would be required to explain the observed large orbital 
period derivative, and conclude that much of the transferred mass
must be lost from the system.
Below we show how the orbital period
derivative we measured can be used to constrain the mass transfer in
the system, and how this strongly indicates that a large fraction of
the mass which the companion tries to transfer to the neutron star is lost 
by the system.

The mass-loss rate from the secondary can be easily calculated as a 
function of the orbital period of the system and the measured orbital 
period derivative
combining the third Kepler law, which must be always satisfied by the
orbital parameters of the system, with 
the condition that in this persistent system the neutron star is accreting mass 
through Roche Lobe overflow. This means that the radius of the secondary follows 
the evolution of the secondary Roche Lobe radius: 
$\dot R_{L2} / R_{L2} = \dot R_2 / R_2$, where for the secondary 
we adopted a mass-radius relation $R_2 \propto M_2^n$ and for the radius 
of the secondary Roche Lobe we adopted the \citet{Paczynski_71}
approximation $R_{L2} = 2/3^{4/3} [q/(1+q)]^{1/3} a$, where $a$ is the 
orbital separation, which is valid for small mass ratios, 
$q = M_2 / M_1 \le 0.8$.
From these conditions it is possible to derive a relation between 
the mass-loss rate from the secondary and the orbital period derivative

\begin{equation}
\label{eq:dotM} 
\dot m_{-8} = 3.5 \times (3n-1)^{-1} m_{2, \; 0.1} \left( 
\frac{\dot P_{-10}}{P_{\rm 5h}}\right),   
\end{equation}
where $\dot m_{-8}$ is the secondary mass derivative (negative since
the secondary star looses mass) in units of $10^{-8} M_\odot \; {\rm yr}^{-1}$, 
and $\dot P_{-10}$ is the orbital period
derivative in units of $10^{-10}$. 
We stress that in Eq.~(\ref{eq:dotM}) 
an expression for the angular momentum losses 
mechanism that drives the evolution of the system (e.g. MB or GR) 
does not explicitly appear.
This is quite relevant because at present there is no general consensus 
on the absolute strength of the MB term nor on its functional dependence on the 
other orbital parameters. 
Indeed the effects of the driving mechanism are implicitly considered through
the orbital period derivative, which is a measured quantity in our case.  

Equation (\ref{eq:dotM}) can be inverted to derive the mass-transfer timescale
$\tau_{\dot M} = M_2 / (- \dot M_2)$ 
\begin{equation}
\label{eq:taumdot} 
\tau_{\dot M} = 2.86 \times (1 - 3n) (P_{\rm 5h} / \dot P_{-10}) \times 
{\rm 10^6 \; yr}.
\end{equation}
On this short time-scale the response of the secondary star must be 
adiabatic.
For $m_{2} \sim 0.3$ (see below) the envelope is convective and the 
appropriate index is $n = -1/3$, in agreement with the condition 
$n < 1/3$ discussed above. With this value we find 
$\tau_{\dot M} \sim 4 \times {\rm 10^6 \; yr}$ for {\adc}. 
We note that the Eddington limit 
(in units of $10^{-8} M_\odot \; {\rm yr}^{-1}$)  
for accretion onto a neutron star is 
$\dot m_{\rm E \; -8} = 1.54 \; R_6(m_1)$, where $R_6(m_1)$ is the neutron star 
radius in units of $10^{6}$ cm, which slightly depends on the neutron star mass
once an equation of state (EoS) for the ultradense matter is adopted. 

Thus, adopting $m_{2, \; 0.1} \ge 3.3$ \citep{Jonker_03} and $n = -1/3$ 
in Eq.~(\ref{eq:dotM}), we have to conclude that the secondary mass 
loss rate in {\adc} is super-Eddington. 
We are therefore forced to conclude that the evolution of the system is highly
non-conservative.

In order to search for a possible evolutionary scenario for {\adc} we make
the assumption 
that the neutron star is accreting at the maximum possible rate, 
{\it i.e.} the Eddington limit. 
It has to be noted that the Eddington limit strictly holds for a 
spherical geometry, and may not be a constraint for highly magnetised
neutron stars for which the accreting matter is channeled onto the magnetic 
polar caps and the geometry of the matter distribution over the Alfv\'en 
surface may not be symmetric (see e.g.\ \citet{Basko_76}). However, 
our assumption is justified because the luminosity
function for highly magnetized neutron stars (usually found in High Mass 
X-ray Binaries) does not disagree with this assumption (see e.g.
\citet{Grimm_02}). In particular, no
highly magnetized neutron star is known to accrete at a rate much higher 
than the Eddington limit, and the most luminous high mass X-ray binaries
containing a neutron star in our Galaxy reach luminosities of the order
of the Eddington limit. Moreover, the 
extrapolated X-ray luminosity of {\adc} does not indicate an extremely
high X-ray luminosity. Hence we do not have any evidence that the 
limiting mass accretion rate in this source is very different from the 
Eddington limit. 

\begin{figure}
\includegraphics[width=9.0cm]{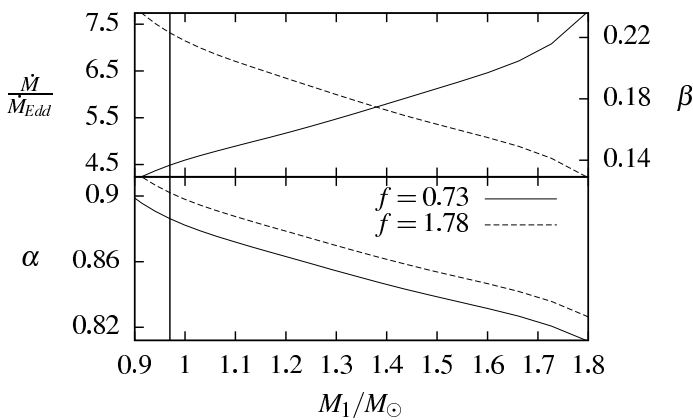}
\caption{Secondary mass loss rate in units of the Eddington limit
for the FPS EoS (top panel, solid line), fraction $\beta$ of the mass 
lost by the secondary star which is accreted onto the neutron star 
(top panel, dashed line),
and specific angular momentum of the mass leaving the system,
$\alpha$, in units of the specific angular momentum of the secondary
computed for two different prescription of the magnetic
braking strength, namely $f = 0.73$ \citep{Skumanich_72} and $f = 1.78$ 
\citep{Smith_79} (bottom panel). 
The vertical line at $0.97\; M_\odot$ represents the lower limit of the 
neutron star mass derived by \citet{Jonker_03}.}
\label{fig:m2dot_and_beta}
\end{figure}

This results in the following condition:
\begin{equation}
\label{eq:dotMacc} 
- \beta \times \dot m_{-8} = 1.54 \; R_6(m_1),  
\end{equation}
where $\beta$ is the fraction of the mass lost by the secondary, which is 
accreted by the neutron star, namely $\dot M_1 = - \beta \dot M_2$, where
$\dot M_1$ and $\dot M_2$ are the mass derivatives of the primary and the 
secondary, respectively.
We consider two EoS, namely the moderately soft FPS and the
stiffer L \citep{Cook_94}, which give the relation $R_6(m_1)$ for the neutron star 
radius.   
With this and considering the mass function of the system derived from the
timing analysis of the neutron star spin \citep{Jonker_01}, 
Eqs.~(\ref{eq:dotM}) and (\ref{eq:dotMacc})
can be solved to derive $\dot M_2$ and $\beta$ for any value of $m_1$ 
between $m_{\rm 1 \; MIN} = 0.97$ and $m_{\rm 1 \; MAX}$, which depends on 
the particular EoS adopted, which is $1.8$ and $2.5$
for EoS FPS and L, respectively.
In Fig.~\ref{fig:m2dot_and_beta} $\dot M_2$, in units of the Eddington 
mass transfer rate, and $\beta$ are plotted for the appropriate range of 
neutron star masses for the FPS case (for the L EoS the value of $\beta$ 
is 30\% higher, while $\dot M_2$ in units of the Eddington limit is
30\% lower). 

We now consider Eq.~(3) of \citet{Disalvo_08}, which expresses    
the conservation of the angular momentum of the system 
\begin{equation}
\label{eq:angmom}
\frac{\dot P_{\rm orb}}{P_{\rm orb}} = 3 \left[\frac{\dot J}{J_{\rm orb}} - 
\frac{\dot M_2}{M_2} \; g(\beta,q,\alpha)\right],
\end{equation}
where, in this case,
\begin{equation}
\label{eq:Jdot/J}
\frac{\dot J}{J_{\rm orb}} = - 5.5 \times 10^{-19} [1.0 + {\rm T_{MB}}]
 \; m_1 \; m_{2, \; 0.1} \; m^{-1/3} 
P_{5h}^{-8/3} 
\end{equation}
represents all the possible losses of angular momentum from 
the system caused by MB and GR, where ${\rm T_{MB}}$ is given
by Eq.~\ref{eq:MB}
and 
\begin{equation}
\label{eq:g}
g(\beta,q,\alpha) = 1 - \beta q - (1-\beta) (\alpha + q/3)/(1+q)
\end{equation}
takes into account the effects of angular momentum losses because of mass loss
from the system. 
$\alpha$ is the specific angular momentum of the mass leaving the system,
$l_{\rm ej}$, in units of the specific angular momentum of the secondary, 
that is: $\alpha = l_{ej} / (\Omega_{\rm orb} r_2^2) = 
l_{ej} P_{orb} (M_1 + M_2)^2/(2 \pi a^2 M_1^2)$, where $r_2$ is the distance 
of the secondary star from the center of mass of the system, and $a$ is the 
orbital separation.

Adopting the two values of $f$ discussed 
above, namely $f = 0.73$ \citep{Skumanich_72} or $f = 1.78$ \citep{Smith_79}, 
and $k_{0.277}=1$, Eqs.~(\ref{eq:MB}), (\ref{eq:angmom}), (\ref{eq:Jdot/J}), 
and (\ref{eq:g}) can be solved to derive $\alpha$ as a function of $m_1$.
In Fig.~\ref{fig:m2dot_and_beta} $\alpha$ is plotted for the appropriate 
range of neutron star masses for the FPS case (for the L EoS the values of 
$\alpha$ are 7\% higher).

The values of $\alpha$ we obtain are in between the specific angular momentum
at the inner Lagrangian point, 
$\alpha_{\rm L1} = [1-0.462(1+q)^{2/3}q^{1/3}]^{2}\sim 0.4$ for {\adc}, and
the specific angular momentum of the secondary, $\alpha_2 = 1.0$, and 
actually quite close to $\alpha_2$. This is expected if the mass lost by the 
secondary star is blown away because of the radiation pressure exerted
by the Eddington luminosity generated by the accretion onto the neutron star.

For both the adopted EoS and all the possible values of the neutron star mass,
the values of $\beta$ are in the range $0.13 - 0.29$, which means that the 
mass transfer in {\adc} is not conservative, at least, at 70\% level, which, 
as we already noted, is true {\it independently of any assumption on the 
particular angular momentum losses} . 
Interestingly, this is the key that opens the possibility
of constructing a consistent secular evolution for this system. Indeed the 
contact condition, $\dot R_{L2} / R_{L2} = \dot R_2 / R_2$, can be solved to 
derive a theoretical prediction for the mass-loss rate once a prescription
is given for the possible losses of angular momentum from the system caused 
by MB and GR as in Eq.~(\ref{eq:Jdot/J}):
$$\dot m_{-8} = - 3.5 \times 10^{-4} [1.0 + {\rm T_{MB}}]
 \; m_1 \; m_{2, \; 0.1}^2 \; m^{-1/3} P_{5h}^{-8/3} \times$$
\begin{equation}
\label{eq:mdotth}
{\rm F}(n,g(\beta,q,\alpha)),   
\end{equation}
where 
\begin{equation}
\label{eq:F}
{\rm F}(n,g(\beta,q,\alpha)) = [n - 1/3 + 2g(\beta,q,\alpha)]^{-1}.
\end{equation}  
The function ${\rm F}(n,g(\beta,q,\alpha))$ is very sensitive to the
scenario adopted: for a conservative scenario ($\beta = 1$)
${\rm F}(n,g(\beta,q,\alpha)) \sim 1.5 $ while for $\beta$ in the range 
0.13 -- 0.22 (which is appropriate for the FPS EoS and all the 
possible values of the neutron star mass) 
${\rm F}(n,g(\beta,q,\alpha)) \sim 40$.
This means that the term acting to shrink the secondary Roche Lobe 
-- because of the extra angular momentum losses caused by the mass expelled 
from the system -- determines an amplification of the mass-loss rate
through the function ${\rm F}(n,g(\beta,q,\alpha))$ with respect to a fully
conservative case. This amplified mass-loss rate in turn determines the 
high value of the orbital period derivative observed in this system. 
Consequential angular momentum losses (CAML), i.e.\ angular momentum losses
that are themselves the result of mass transfer, have been proposed in
the context of cataclysmic variables (CVs) evolution (see e.g.\
\citet{King_95}).


Inserting the values determined in this paper for {\adc} in 
Eq.~(\ref{eq:taumdot}) we find $\tau_{\dot M} = 2.1 \times {\rm 10^6 \; yr}$.
This means that the system as it is observed now will probably end on 
this timescale, possibly with the tidal disruption of the companion star. 
Indeed, \citet{King_95} argued that mass transfer 
could be unstable when CAML are present.
This time-scale is extremely short, which indicates that it is possible that 
some short orbital period LMXBs can last much shorter than what was previously 
thought. This evolutionary phase, characterised by a super-Eddington 
mass transfer rate, may be a common phase in the evolution of LMXBs, 
albeit short-living. Because this phase should not last more than a 
few million years, there may be very few observed systems in this phase 
(e.g. the so-called Z-sources, which are persistently bright LMXBs). 
This could have profound implications for the estimate of
the actual number of LMXBs produced in the Galaxy as inferred from the 
observed ones, and also for 
the predicted number of millisecond binary pulsar. We note that this
would help to bring the number of LMXBs in line with the estimated number
of millisecond binary pulsars. But a detailed analysis of this delicate
and long-standing problem needs a dedicated study of this almost unstable phase
of the orbital evolution, which is beyond the scope of this paper and will 
be discussed in a forthcoming paper.

\begin{acknowledgements}
This work is supported by the Italian Space Agency,
ASI-INAF I/088/06/0 contract for High Energy Astrophysics. 
We thank the referee for useful comments on the manuscript.
\end{acknowledgements}

\bibliographystyle{aa}
\bibliography{12881}

\end{document}